\documentclass[aps,prl,preprint,superscriptaddress]{revtex4-1}

\usepackage{graphicx}

\begin{document}
\title{Shercliff layers in strongly magnetic cylindrical\\ Taylor-Couette flow}

\author{Rainer Hollerbach$^1$, Deborah Hulot}
\affiliation{Department of Applied Mathematics, University of Leeds,
Leeds LS2 9JT, UK}
\affiliation{Institut National des Sciences Appliqu\'ees de Rouen,
76801 Saint Etienne du Rouvray Cedex, France}

\date{\today}

\begin{abstract}
We numerically compute axisymmetric Taylor-Couette flow in the presence
of axially periodic magnetic fields, with Hartmann numbers up to $Ha^2=10^7$.
The geometry of the field singles out special field lines on which Shercliff
layers form.  These are simple shear layers for insulating boundaries, versus
super-rotating or counter-rotating layers for conducting boundaries.  Some
field configurations have previously studied spherical analogs, but
fundamentally new configurations also exist, having no spherical analogs.
Finally, we explore the influence of azimuthal fields $B_\phi\sim
r^{-1}{\bf\hat e}_\phi$ on these layers, and show that the flow is
suppressed for conducting boundaries but enhanced for insulating boundaries.\\
$\phantom{x}\qquad\qquad\qquad\qquad\qquad\qquad\qquad\qquad\qquad\;$
{\large R\'esum\'e}\\
Nous mod\'eliserons l'\'ecoulement axisym\'etrique de Taylor-Couette en
pr\'esence d'un champ magn\'etique axialement p\'eriodique, avec un nombre
de Hartmann jusqu'\`a $Ha^2=10^7$. La g\'eometrie du champ montre des lignes
de champ sous forme de couche de Shercliff. Il y a des couches de cisaillement,
lorsque les fronti\`eres sont isolantes, tandis que la rotation est excessive
ou invers\'ee pour les fronti\`eres conductrices. Certaines configurations de
champs sont similaires \`a celles vues sous forme sph\'erique cependant de
nouvelles configurations existent. Enfin, nous d\'ecouvrirons l'influence de
champs azimutaux ($B_\phi\sim r^{-1}{\bf\hat e}_\phi$) sur ces couches et nous
montrerons que l'\'ecoulement diminue avec des bords conducteurs alors qu'il
s'accentue pour des fronti\`eres isolantes.
\end{abstract}

\maketitle

\section{Introduction}

Shercliff layers are free shear layers that can occur in the flow of an
electrically conducting fluid when a sufficiently strong magnetic field
is externally imposed \cite{Shercliff}.  They arise due to the strongly
anisotropic nature of the Lorentz force, consisting of a tension along the
magnetic field lines.  The details of how the spatial structure of the
imposed field overlaps with the geometry of the container can then single
out special field lines on which Shercliff layers form.

For example, suppose we consider spherical Couette flow, the flow induced in
a spherical shell where the inner sphere rotates and the outer one is fixed.
Consider further two possible choices of magnetic fields to impose, a dipole
${\bf B}_d=2\sigma^{-3}\cos\theta\,{\bf\hat e}_\sigma
+\sigma^{-3}\sin\theta\,{\bf\hat e}_\theta$ and a uniform axial field
${\bf B}_a={\bf\hat e}_z=\cos\theta\,{\bf\hat e}_\sigma-\sin\theta\,
{\bf\hat e}_\theta$, where $(\sigma,\theta,\phi)$ are standard spherical
coordinates, and $(z,r,\phi)$ cylindrical coordinates.
For the dipole field, there will be some field lines that link only to the
inner sphere, and others that connect the two spheres.  Similarly, for the
axial field there will be some field lines that link only to the outer sphere,
and others that connect the two spheres.  The tension in the field lines then
ensures that any field lines linked to one boundary only are completely locked
to that boundary, with the fluid either co-rotating with the inner sphere, or
stationary together with the outer sphere.  It is only on field lines that
connect to both boundaries that the fluid is faced with conflicting conditions
at the two ends of the line, and resolves this conflict by rotating at a rate
intermediate between the two end values.

The entire domain is therefore naturally divided up into different regions
depending on how the field lines connect to the boundaries, with the angular
velocity changing abruptly across those field lines separating different
regions \cite{Dormy1,Starchenko}.  Furthermore, it is clear that there is
nothing special about either the spherical geometry or these two particular
fields.  As long as both the container and the imposed field are axisymmetric,
the same considerations will apply, and will always result in Shercliff layers
forming on these special field lines where the linkage to the boundaries
switches from one type to another.  The thickness of these layers scales as
$Ha^{-1/2}$, where the Hartmann number $Ha$ is a measure of the strength of
the imposed field \cite{Roberts}.

Another intriguing result is the influence of the electromagnetic boundary
conditions.  The conclusion above, that Shercliff layers are simply shear
layers on which the angular velocity switches to something intermediate
between 0 at the outer boundary and 1 at the inner boundary, is valid only if
both boundaries are insulating.  If instead the inner sphere is conducting,
a dipole field yields a so-called super-rotation, where the fluid within
the Shercliff layer rotates faster than the inner sphere \cite{Dormy1}.
Alternatively, if the outer sphere is conducting, an axial field yields a
counter-rotation, where the fluid within the Shercliff layer rotates in the
opposite direction to the inner sphere \cite{Hollerbach00}.  In both of
these cases, the degree of super-rotation or counter-rotation is around
20-30\% of the inner sphere's rotation rate, independent of $Ha$ (for
sufficiently large values).  Even more unexpected results are obtained if
both boundaries are taken to be conducting; in this case the degree of
`anomalous' rotation appears to increase indefinitely as $Ha$ is increased in
a numerical computation \cite{Hollerbach00,Hollerbach01}.  Various asymptotic
analyses of this problem confirm that the anomalous rotation should be $O(1)$
if only one boundary is conducting, but $O(Ha^{1/2})$ if both boundaries are
conducting \cite{Dormy2,Mizerski,Buhler,Soward}.

Motivated by these counter-intuitive results, \cite{Hollerbach01b}
performed a systematic investigation of linear combinations of dipole and
axial fields, and showed that it is even possible to obtain both super-rotation
and counter-rotation simultaneously.  One finds easily enough that combinations
of these two basic ingredients, dipole and axial, are sufficient to create all
field line topologies that are possible in a spherical shell geometry.  The
purpose of this paper is to show that other topologies are possible in
cylindrical geometry, and to numerically explore what happens in those cases.
For example, we will show that it is possible to construct a field having a
single field line that is tangent to both the inner and outer cylinders, with
the tangency at the outer cylinder then suggesting a super-rotation, but the
tangency at the inner cylinder suggesting a counter-rotation.  So what does
happen in that case?  We will further explore what happens when azimuthal
fields of the form $r^{-1}{\bf\hat e}_\phi$ are added, which also have no
natural analog in spherical geometry.

Finally, it is worth noting that there have been several liquid metal
experiments related to some of the topics considered here.  These include
spherical Couette flow in both dipole \cite{Nataf1,Brito,Cabanes} and axial
\cite{Sisan,Zimmerman} fields, cylindrical Taylor-Couette flow in an axial
field \cite{Spence,Roach}, and even electromagnetically driven flows
\cite{Stelzer1,Stelzer2}.  However, inertia (finite Reynolds number) plays
an important role in most of these results, unlike in the `pure' Shercliff
layer problem considered here.  See also
\cite{Hollerbach07,Hollerbach09,Gissinger1,Gissinger2,Figueroa,Nataf2,Kaplan}
for numerical results related to some of these experiments, as well as
\cite{Rudiger} for a general review of magnetohydrodynamic Couette flows.

\section{Equations}

We consider a cylindrical Taylor-Couette geometry with nondimensional radii
$r_i=1$ and $r_o=2$.  Periodicity is imposed in $z$, with a wavelength
$z_0=4$.  The precise choice $z_0=4$ is not crucial, with a
broad range of $O(1)$ values yielding similar Shercliff layer structures.
(Taking $z_0\gg O(1)$ could well lead to different solutions though.)

In the inductionless limit, the nondimensional Navier-Stokes and
magnetic induction equations are
$$\frac{\partial\bf U}{\partial t}=-\nabla p + \nabla^2{\bf U}
  - Re{\bf U\cdot\nabla U} + Ha^2(\nabla\times{\bf b})\times{\bf B}_0,
\eqno(1)$$
$$\nabla^2{\bf b}=-\nabla\times({\bf U\times B}_0),\eqno(2)$$
where $\bf U$ is the fluid flow, ${\bf B}_0$ is the externally imposed
magnetic field, and $\bf b$ the induced field.  For the axisymmetric
solutions that are relevant here, it is convenient to further decompose
$\bf U$ and $\bf b$ as
$${\bf U}=\nabla\times(\psi\,{\bf\hat e}_\phi) + v\,{\bf\hat e}_\phi,
\qquad\qquad
  {\bf b}=\nabla\times(a\,{\bf\hat e}_\phi) + b\,{\bf\hat e}_\phi.\eqno(3)$$

The two nondimensional parameters are the Hartmann number
$Ha={B_0 r_i}/{\sqrt{\mu\rho\nu\eta}}$
measuring the strength $B_0$ of the imposed field, and the Reynolds number
$Re={\Omega r_i^2}/{\nu}$
measuring the inner cylinder's rotation rate $\Omega$.  In fact, in this work
we are interested in the limit of infinitesimal differential rotation, so we
set $Re\to0$ and remove the inertial term $Re{\bf U\cdot\nabla U}$, but again
see \cite{Nataf1,Brito,Cabanes,Sisan,Zimmerman,Spence,Roach,Stelzer1,Stelzer2,
Hollerbach07,Hollerbach09,Gissinger1,Gissinger2,Figueroa,Nataf2,Kaplan} for a
broad variety of effects that can arise at finite $Re$.  The quantities $\mu$,
$\rho$, $\nu$, and $\eta$ are the fluid's permeability, density, viscosity,
and magnetic diffusivity, respectively.

We next turn to the allowed choices for the imposed field ${\bf B}_0$.  The
requirements are that it should be axisymmetric, periodic in $z$, and satisfy
$\nabla\cdot{\bf B}_0=0$ and $\nabla\times{\bf B}_0={\bf0}$.  The condition
$\nabla\cdot{\bf B}_0=0$ is of course one of Maxwell's original equations;
the condition $\nabla\times{\bf B}_0={\bf0}$ states that ${\bf B}_0$ is a
potential field, and not due to electric currents within the fluid.  In
addition to the familiar $z$-independent fields ${\bf\hat e}_z$ and
$r^{-1}{\bf\hat e}_\phi$, the only other choices that satisfy all four of
these conditions are
$${\bf B}_I =\cos(\kappa z)I_0(\kappa r)\,{\bf\hat e}_z
           + \sin(\kappa z)I_1(\kappa r)\,{\bf\hat e}_r,\eqno(4)$$
$${\bf B}_K =\cos(\kappa z)K_0(\kappa r)\,{\bf\hat e}_z
           - \sin(\kappa z)K_1(\kappa r)\,{\bf\hat e}_r,\eqno(5)$$
where $\kappa=2\pi/z_0$ and $I_0$, $I_1$, $K_0$ and $K_1$ are the modified
Bessel functions \cite{AandS}.  ${\bf B}_I$ corresponds to the field that
would be generated by an array of Helmholtz coils in the region $r>r_o$,
with currents that alternate periodically in $z$; ${\bf B}_K$ is the field
that would be generated by squeezing the Helmholtz array into the region
$r<r_i$.

\begin{figure}
  \centerline{\includegraphics{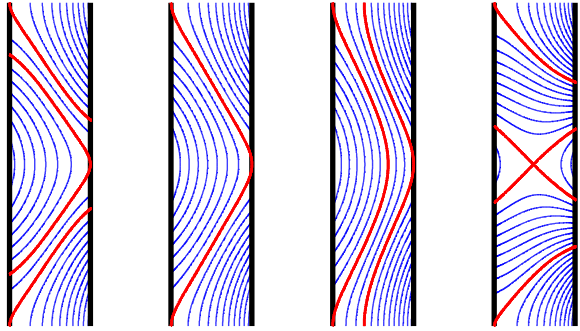}}
  \caption{(Color online.) From left to right, the four combinations that
are referred to as Fields 1-4 in the text.  The thick red lines in each
case denote the special field lines across which the linkage to the boundaries
changes, and where Shercliff layers are thus expected to arise.  The general
form is $({\bf\hat e}_z + c_I{\bf B}_I + c_K{\bf B}_K)/c_0$, where $c_I$ and
$c_K$ are adjusted to select the desired field topology, and $c_0$ is then
chosen to rescale the maximum amplitude to 1.  The precise values of $(c_I,
c_K)$ are $(0.155,-4.356)$, $(0.124,-3.485)$, $(0.091,-2.562)$,
$(0.548,-7.687)$, respectively.}
\label{fig1}
\end{figure}

Various linear combinations of ${\bf\hat e}_z$, ${\bf B}_I$ and ${\bf B}_K$
then correspond to the linear combinations of dipole and axial fields
discussed above.  One easily finds that it is possible to construct field
line topologies that have no analog in the spherical shell geometry.
Fig.\ 1 shows four possible combinations.
In Field 1, there are some field lines that thread only the inner
cylinder, some that thread only the outer cylinder, and some that connect the
two.  Based on the spherical results, if both boundaries are conducting, we
would expect to find a super-rotating jet on one dividing line, and a
counter-rotating jet on the other.
In Field 2, the linear combinations have been adjusted slightly, in
such a way that the previously separate dividing field lines coincide, and
the region of field lines linking both boundaries has been collapsed to
this single line that is tangent to both boundaries, but does not penetrate
either.  So, what happens in this case, when the previously expected
super-rotating and counter-rotating jets should now occur on one and the
same field line?
In Field 3, the linear combinations have been further adjusted, so
there is now a central ribbon of field lines that never touch either
boundary, but just continue periodically to $z=\pm\infty$.  As with
Field 2, this scenario also has no spherical analog, so it is again
not clear what to expect in this case.
Finally, in Field 4 the linear combinations have been chosen to yield
an X-type neutral point in the middle of the domain.  This topology is in
fact achievable in spherical geometry as well \cite{Hollerbach01b}, so is
included here primarily for completeness and comparison.

We see then that just taking different combinations of  ${\bf\hat e}_z$,
${\bf B}_I$ and ${\bf B}_K$ already allows us to construct topologies that
have no spherical analogs.  To all of these we can further add the
azimuthal field $r^{-1}{\bf\hat e}_\phi$, which also has no natural
analog in spherical geometry, since it is singular on the $z$-axis, which
is part of the domain in spherical geometry but not in cylindrical.  Since
it is everywhere tangent to the boundaries, this azimuthal field will not
alter the fundamental topology of the previously considered fields, but it
nevertheless changes the detailed structure of the Shercliff layers that
arise on the critical field lines.  Indeed, including an azimuthal field
component changes the solutions in at least one quite fundamental way:
For purely meridional fields, the coupling between the different quantities
turns out to be such that in fact $\psi$ and $a$ in Eq.\ (3) are identically
zero (in the $Re\to0$ limit).  Adding an azimuthal component to ${\bf B}_0$
introduces new couplings that result in non-zero $\psi$ and $a$.
Finally, note also that a {\it purely} azimuthal ${\bf B}_0$
would not yield any interesting dynamics; the solution in that case is
simply the original Couette profile $v=(-r+4r^{-1})/3$, and $\psi=a=b=0$.

To summarize, the goal of this paper is to explore the Shercliff layers
that occur on the critical field lines indicated in Fig.\ 1, either these
fields alone or together with azimuthal fields of the form
$r^{-1}{\bf\hat e}_\phi$.  This is accomplished by using an axisymmetric,
pseudo-spectral code \cite{Hollerbach08} to numerically solve Eqs.\ (1)-(3).
Very briefly, $\psi$, $v$, $a$ and $b$ are expanded in
terms of Chebyshev polynomials in $r$ and Fourier series in $z$.  Eq.\ (1) is
time-stepped until a stationary solution emerges; Eq.\ (2) is directly
inverted for $\bf b$ at each time-step of Eq.\ (1).  Resolutions as large as
240 Chebyshev polynomials and 400 Fourier modes were used, and allow Hartmann
numbers as large as $Ha^2=10^7$ to be achieved.

The associated boundary conditions are no-slip for $\bf U$, and either
insulating or perfectly conducting boundaries for $\bf b$, referred to as I
and C respectively.  Other possible choices could include
finitely conducting, or perhaps ferromagnetic, which in other contexts can
have a significant influence \cite{Gissinger3}.  For the Shercliff layer
problem the asymptotic analyses \cite{Dormy2,Mizerski,Buhler,Soward}
indicate that the most relevant parameter is how the conductance of the
exterior regions compares with the conductance of the fluid region; if this
ratio is small (large) the results are similar to the insulating (perfectly
conducting) case.  Our I and C choices are therefore natural limiting cases,
and even something at first sight quite different, such as ferromagnetic, is
likely to be similar to the I case, since they both have zero conductance of
the exterior regions.

\section{Results without imposed $B_\phi$}

Fig.\ 2 shows contours of the angular velocity $\omega=v/r$ for the four
choices Fields 1-4 alone, without any additional azimuthal component.
In every case, the most prominent features are indeed concentrated on the
particular field lines singled out in Fig.\ 1.  The contrast between
insulating and conducting boundaries is also clear; conducting boundaries
exhibit both super-rotation and counter-rotation, especially for Field 4,
whereas insulating boundaries only have very weak counter-rotating regions.

\begin{figure}
  \centerline{\includegraphics{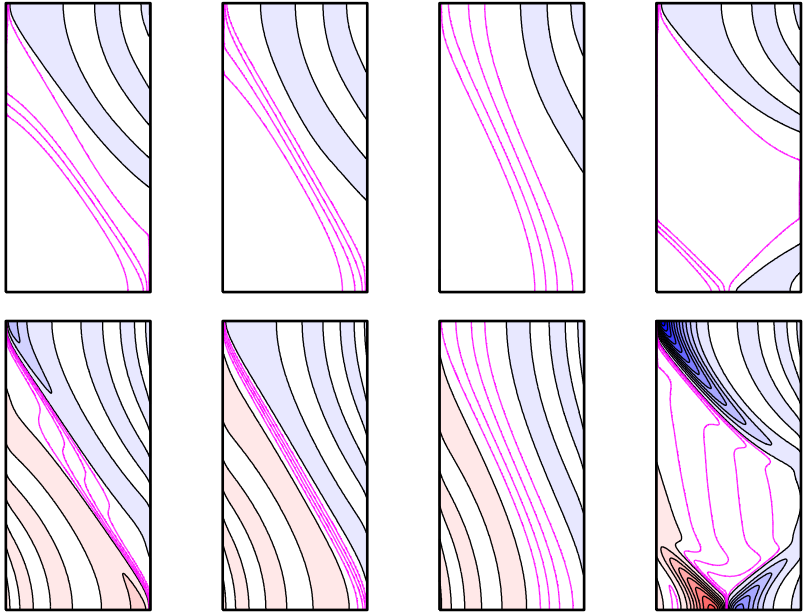}}
  \caption{(Color online.) Contours of the angular velocity $\omega=v/r$,
for $Ha^2=10^6$.  From left to right are the four choices Fields 1-4.
The top row is for insulating boundaries, the bottom row conducting.  In
each panel only the upper half of the domain is shown; that is, $r$ extends
over the full range $[1,2]$, but $z\in[2,4]$.  The lower half $z\in[0,2]$
is reflection-symmetric in each case, as seen also in Fig.\ 1.  The white
regions indicate values between 0 at the outer boundary and 1 at the inner;
the magenta contour lines in these regions have intervals 0.2.  The
red-shaded regions correspond to super-rotation, where $\omega>1$; the
blue-shaded regions correspond to counter-rotation, where $\omega<0$.  In
both cases the black contour lines in these regions have intervals 1.}
\label{fig2}
\end{figure}

In the insulating case there are also Hartmann layers at the
boundaries.  These layers are so thin, $O(Ha^{-1})$, that they cannot be seen
directly at this scale.  Their presence can be inferred though by the magenta
contour lines, indicating values between 0.2 and 0.8, that appear to touch
the boundaries.  The actual imposed boundary conditions of course are $\omega
=1$ at $r_i$ and $\omega=0$ at $r_o$, so these contour lines cannot touch the
boundaries, and indeed they don't, but rather remain within the Hartmann
layers.  These layers were investigated in detail, and always followed the
expected $O(Ha^{-1})$ scalings.  We therefore concentrate only on the
Shercliff layers in the following discussion.

To explore the details of the Shercliff layers, we require
more precise diagnostics than the two-dimensional contour plots in Fig.\ 2.
Fig.\ 3 shows one-dimensional cuts along the midplane $z=2$.  Such cuts allow
much more quantitative information to be extracted, such as how the thicknesses
and amplitudes scale with $Ha$.  The thicknesses were again always found to be
broadly consistent with the expected $O(Ha^{-1/2})$ scalings.  Regarding the
amplitudes, insulating boundaries are as expected, with hardly any anomalous
rotation for any of Fields 1-4.

\begin{figure}
  \centerline{\includegraphics{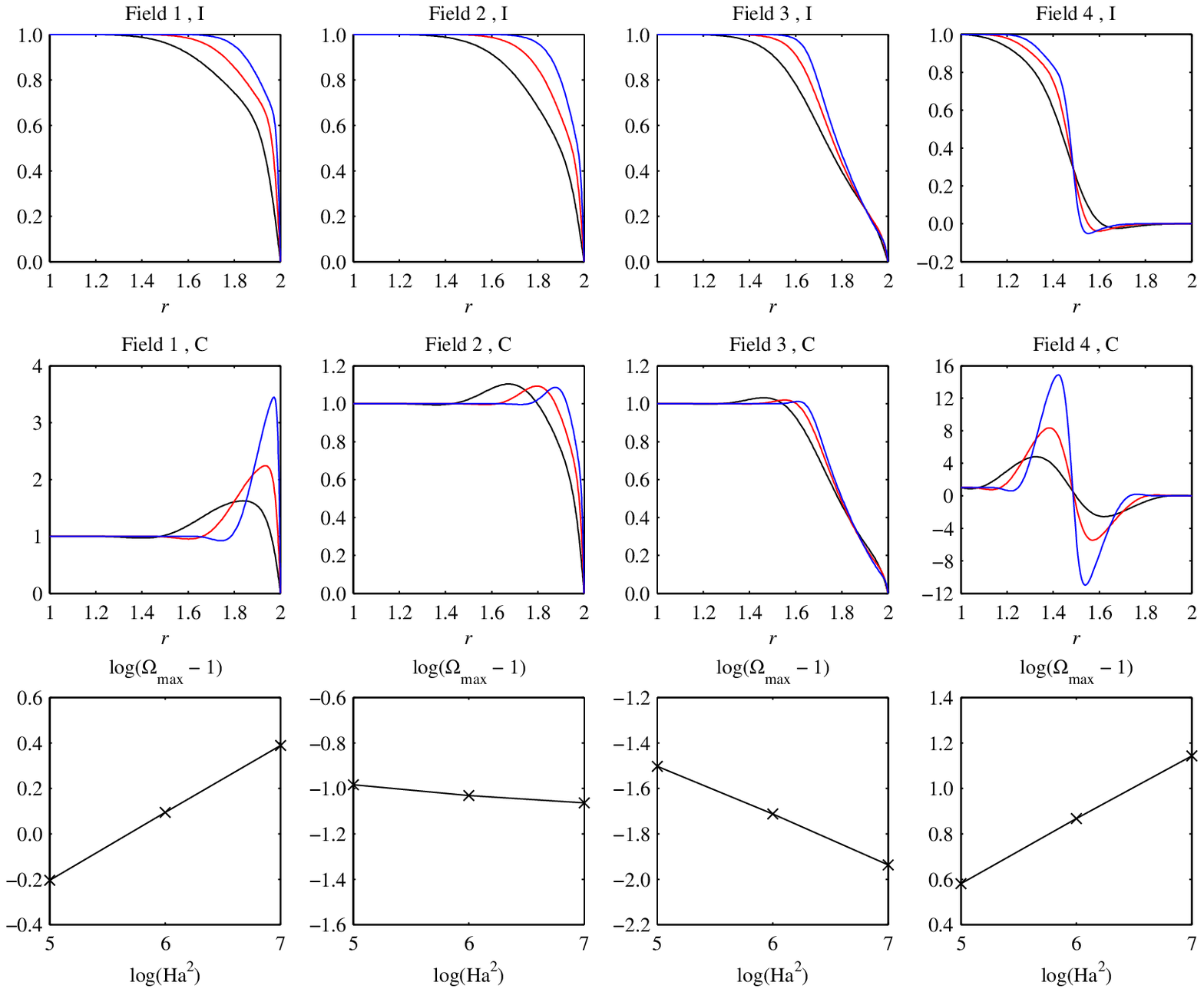}}
  \caption{(Color online.) The first two rows show $\omega(r)$ at $z=2$, for
Fields 1-4 from left to right as indicated, the first row insulating (I) and
the second row conducting (C).  Within each panel black-red-blue indicate
$Ha^2=10^5$, $10^6$, $10^7$, respectively.  The third row shows how the
amplitude of the super-rotation in the second row scales with $Ha$, and
suggests fits of the form $Ha^s$, with $s\approx0.59$, $-0.08$, $-0.43$,
$0.56$, respectively.}
\label{fig3}
\end{figure}

Conducting boundaries exhibit precisely the features we were
expecting, and which make this problem interesting.  Starting with Field 1,
we see that the super-rotating jet on the field line tangent to the outer
boundary is clearly increasing with increasing $Ha$, apparently scaling as
$Ha^{0.59}$.  The counter-rotating jet on the field line tangent to the inner
boundary at $z=4$ has much the same scaling.  Similarly for Field 4, we see
the same behaviour even more strongly, for both the super-rotating and
counter-rotating jets.  Field 4 in particular is not only
qualitatively, but even quantitatively very similar to corresponding results
in spherical geometry -- compare for example with Fig.\ 3 of
\cite{Hollerbach01b}.

In contrast, Field 2 still exhibits a slight super-rotation, but its
amplitude seems to be practically independent of $Ha$.  Similarly, a cut at
$z=4$ has a slight counter-rotation, also with an $Ha$-independent amplitude.
We recall that Field 2 is the case where the previously distinct field lines
in Field 1 have been made to coincide.  Evidently the system adjusts in such
a way that weak anomalous rotations remain, but they no longer increase with
increasing $Ha$.  Finally, for Field 3, having this ribbon of field lines
that are not connected to either boundary gives the system so much flexibility
in adjusting the shear across the Shercliff layers that the anomalous rotation
decreases with increasing $Ha$, apparently scaling as $Ha^{-0.43}$.

To understand the origin of the anomalous rotations, we turn to the Lorentz
force $Ha^2(\nabla\times{\bf b})\times{\bf B}_0$ in Eq.\ (1).  Fig.\ 4 shows
contours of the streamfunction of the induced current $\bf j=\nabla\times b$,
for Fields 1 and 3.  For both choices, I and C boundaries yield very similar
patterns, consisting of clockwise circulation cells.  Focusing attention
specifically at the point $(r,z)=(2,2)$, the current in all four cases is
therefore in the $-{\bf\hat e}_r$ direction.  Since ${\bf B}_0$ at this point
is in the ${\bf\hat e}_z$ direction, the Lorentz force will be in the
${\bf\hat e}_\phi$ direction.  It is precisely this force which accelerates
the fluid from $\omega=0$ at the boundary to $\omega>0$ in the interior.
For insulating boundaries this force is just sufficient to achieve
$\omega\approx1$ on those field lines linked only to the inner boundary,
as seen in Figs.\ 2 and 3.

\begin{figure}
  \centerline{\includegraphics{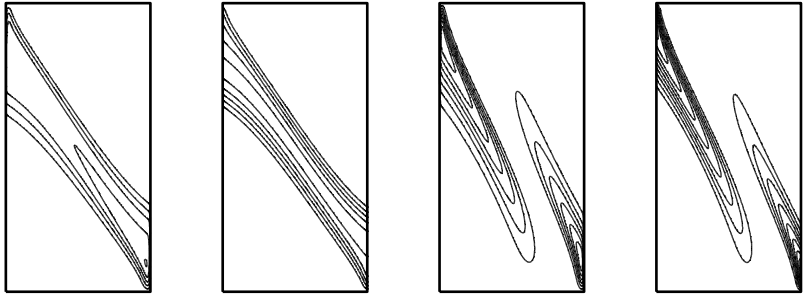}}
  \caption{Contours of $b\,r$, which constitutes the streamfunction of the
electric current ${\bf j}=\nabla\times(b\,{\bf\hat e}_\phi)$.  From left to
right Field 1, I and C boundaries, then Field 3, I and C boundaries, and
$Ha^2=10^6$ for all four.  All circulation cells are clockwise, with
recirculation within the Hartmann boundary layers for the I cases, and
through the boundaries for the C cases.  As in Fig.\ 2, only the upper half
$z\in[2,4]$ is shown; the circulation cells in the lower half are
counter-clockwise.  Finally, the contour intervals from left to right are
$2\cdot10^{-4}$, $2\cdot10^{-3}$, $2\cdot10^{-5}$ and $3\cdot10^{-5}$,
respectively, and illustrates how switching the boundaries from I to C has a
far greater effect for Field 1 than for Field 3.}
\label{fig4}
\end{figure}

For conducting boundaries the system essentially
`over-reacts', and thereby causes the super-rotation in this region.  To
understand further why the system over-reacts in this way, we need to consider
two (closely related) differences between the insulating and conducting results
in Fig.\ 4.  Although the patterns are generally similar for both boundary
conditions, in the insulating case the current must recirculate through the
Hartmann boundary layers (which are again so thin as to be barely visible
here), whereas in the conducting case the current can recirculate through the
exterior regions.  Recirculating the current is therefore much easier in the
conducting case, resulting in a stronger current, hence a stronger Lorentz
force, hence the over-reaction.  As indicated in Fig.\ 4, for Field 1 the
current is an order of magnitude greater for C than for I boundaries,
consistent with the increasing super-rotation, whereas for Field 3 it is only
moderately greater, consistent with much weaker, and indeed decreasing
super-rotation.

The various other anomalous rotations, at other locations, and also for Fields
2 and 4, are similarly explained by the orientation of the Lorentz force at
the position in question.  The results for Fields 1 and 4 are fully consistent
with the analogous scalings previously obtained in the spherical problem
\cite{Hollerbach00,Hollerbach01,Buhler,Soward}.  The new cases, Fields 2 and
3, would certainly also merit further asymptotic analyses to discover the
precise scalings in these cases, and why they differ from the previous results.

\section{Results with imposed $B_\phi$}

To all the cases studied so far, we now wish to add azimuthal fields of
the form $B_\phi=\beta r^{-1}{\bf\hat e}_\phi$, with amplitudes $\beta>0$.
This is again a configuration that has not been considered before, but one
that fundamentally alters the nature of the $Re\to0$ `pure' Shercliff layer
problem.  If ${\bf B}_0$ includes an azimuthal component, then the Lorentz
force $Ha^2(\nabla\times{\bf b})\times{\bf B}_0$ in Eq.\ (1) will include a
meridional component, thereby driving a meridional circulation $\nabla\times
(\psi\,{\bf\hat e}_\phi)$ that would otherwise be absent.  Once $\psi\neq0$,
Eq.\ (2) will similarly induce a field $\nabla\times(a\,{\bf\hat e}_\phi)$.
In the process the previous $v$ and $b$ will also be modified.  We will focus
especially on how the angular velocity is altered, as well as the new flow
component $\nabla\times(\psi\,{\bf\hat e}_\phi)$.  We gradually increased
$\beta$ from 0, and found that $\beta=O(1)$ is already sufficient to
noticeably change the previous results.  However, the most significant
adjustments seem to occur for somewhat larger values, so we fix $\beta=10$
in the following.  (That is, the Hartmann number continues to measure the
strength of the imposed meridional field, but the imposed azimuthal field is
$\sim10$ times stronger.)

\begin{figure}
  \centerline{\includegraphics{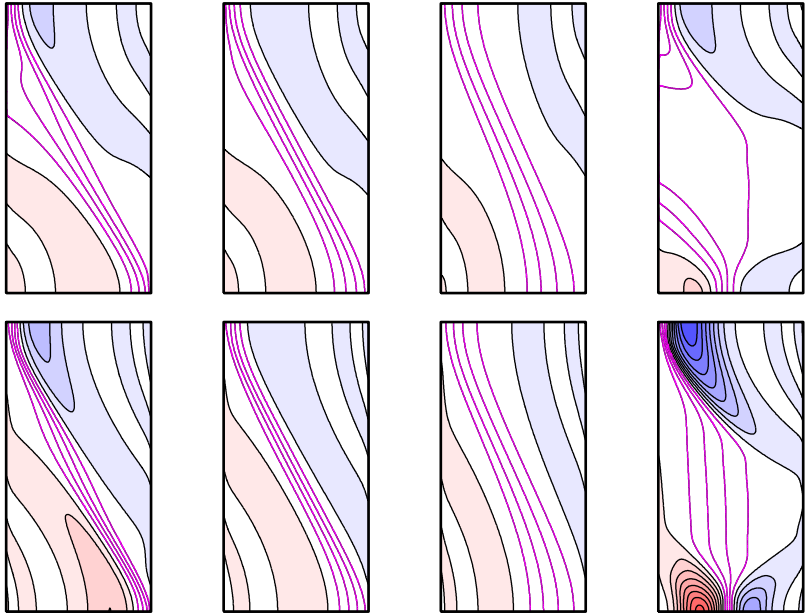}}
  \caption{(Color online.) Contours of the angular velocity $\omega=v/r$,
for $Ha^2=10^6$, and with $B_\phi=10r^{-1}{\bf\hat e}_\phi$ added to the
previous choices Fields 1-4.  All eight panels are exactly as in Fig.\ 2,
except that the contour interval is now 0.2 throughout, for both the magenta
and the black contour lines.}
\label{fig5}
\end{figure}

\begin{figure}
  \centerline{\includegraphics{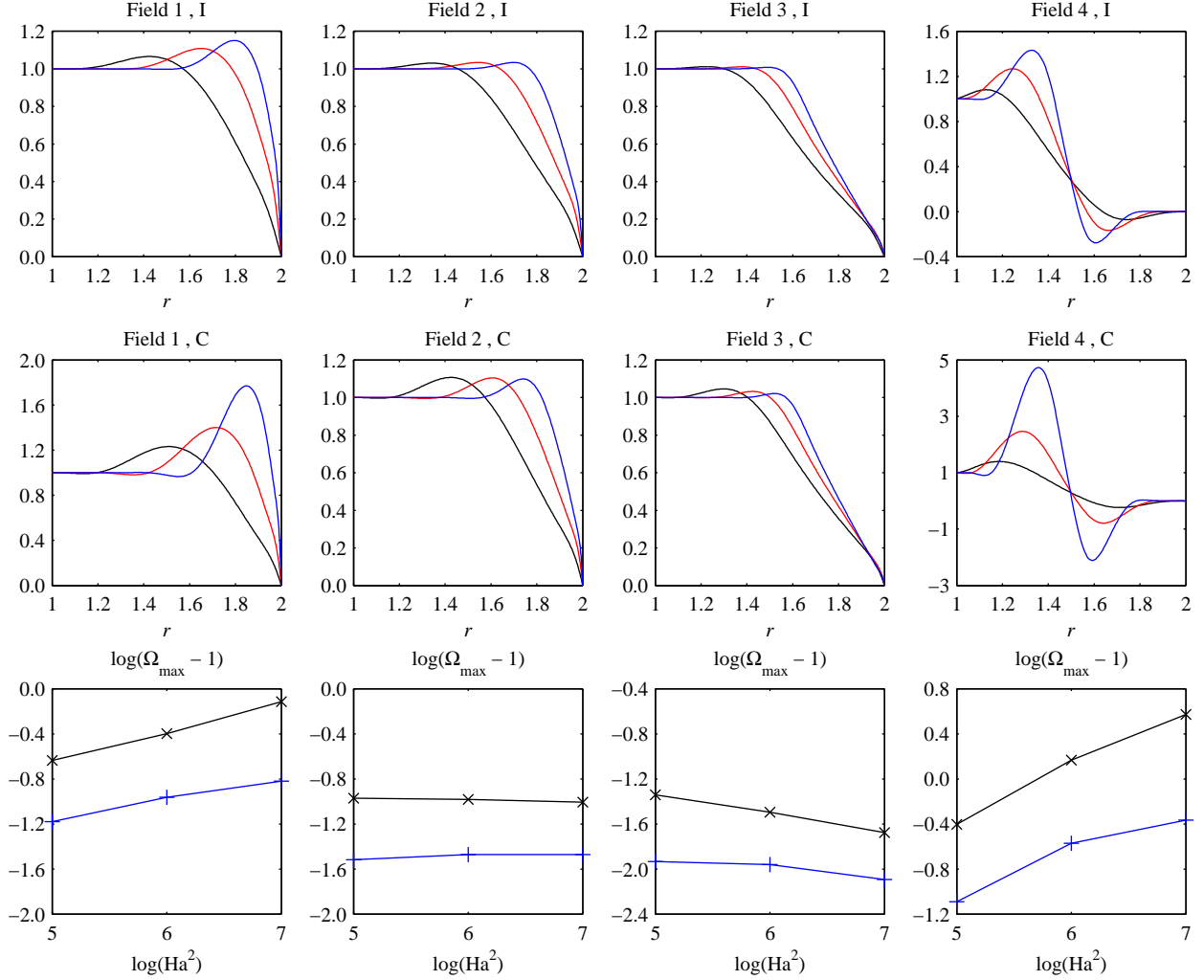}}
  \caption{(Color online.) As in Fig.\ 3, the first two rows show $\omega(r)$
at $z=2$, for Fields 1-4 and I and C as indicated.  Within each panel
black-red-blue again corresponds to $Ha^2=10^5$, $10^6$, $10^7$.  In the third
row, the blue lines ($+$ symbols) and the black lines ($\times$ symbols) show
the scalings with $Ha$ of the I and C super-rotations, respectively.}
\label{fig6}
\end{figure}

\begin{figure}
  \centerline{\includegraphics{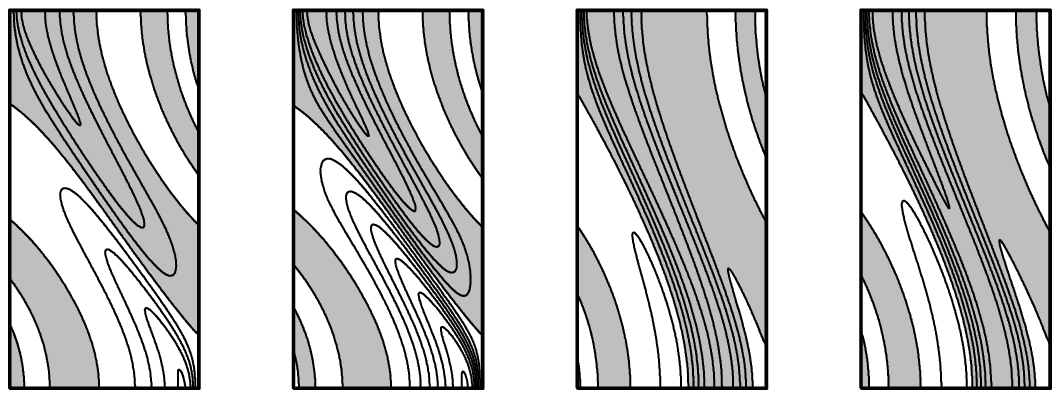}}
  \caption{Contours of $\psi\,r$, which constitutes the streamfunction of the
meridional circulation $\nabla\times(\psi\,{\bf\hat e}_\phi)$.  From left to
right Field 1, I and C boundaries, then Field 3, I and C boundaries, and
$Ha^2=10^6$ for all four.  White indicates negative values, grey positive.
The contour intervals are $5\cdot10^{-3}$ for Field 1, and $10^{-3}$ for
Field 3.}
\label{fig7}
\end{figure}

Fig.\ 5 shows the equivalent of Fig.\ 2.  The qualitative features are still
similar, but there are also clear differences.  Most notably, the very strong
anomalous rotations in the conducting case have been substantially reduced.
All of the various Shercliff layers also seem to be considerably thicker than
before, although an examination of the variation with $Ha$ still suggests a
scaling as $O(Ha^{-1/2})$.

Fig.\ 6 again shows cuts at $z=2$.  Comparing with
Fig.\ 3, the key differences are: (a) the broadening of the Shercliff layers
already noted above, (b) the presence of anomalous rotation in the I case,
(c) the strong suppression of anomalous rotation in the C case, and (d) the
broadly similar scalings of the anomalous rotations in the I and C cases.  We
note though that the anomalous rotation scalings in most cases are not as
clear as in Fig.\ 3; for Fields 1-3 one might conjecture scalings roughly
as $s\approx0.45$, $0.0$ and $-0.3$, respectively, but for Field 4 one
probably should not speculate about a particular exponent at all.

Finally, Fig.\ 7 shows examples of the meridional circulation.  As one might
expect, it also tends to align with the previously existing Shercliff layers,
which continue to dominate the flow, that is, $U_\phi\gg U_z,\,U_r$.  For both
choices of imposed field the I and C options also yield broadly similar
magnitudes of $\psi$.

We conclude this section, and this paper, by noting that while the
$\beta>0$ case yields Shercliff layers similar in many ways to the previously
studied $\beta=0$ case, there are also clear differences, and many of the
precise scalings are almost certainly different.  An asymptotic analysis of
this problem along the lines of the previous analyses
\cite{Dormy2,Mizerski,Buhler,Soward} would be of considerable interest.

\section{Acknowledgments}
DH's visit to Leeds was supported by an Erasmus+ scholarship and by
`Region Stages mobilit\'e' from Haute-Normandie.

\end{document}